# Freezing and Melting Hysteresis Measurements in Solutions of Hyperactive Antifreeze Protein from an Antarctic Bacteria

Yeliz Celik[1], Ran Drori[2], Laurie Graham[3], Yee-Foong Mok[3], Peter L. Davies[3], Ido Braslavsky[1,2*]

[1] Department of Physics and Astronomy, Ohio University, Athens, OH, 45701, USA
[2] Institute of Biochemistry, Food Science and Nutrition, Faculty of Agricultural, Food and Environmental Quality Sciences, The Hebrew University of Jerusalem, Rehovot 76100, Israel,
[3] Department of Biochemistry, Queen's University, Kingston, Ontario K7L 3N6, Canada

*Corresponding Author. Email: braslavs@agri.huji.ac.il

Antifreeze proteins (AFPs) evolved in cold-adapted organisms and serve to protect them against freezing in cold conditions by arresting ice crystal growth. Recently, we have shown quantitatively that adsorption of AFPs not only prevents ice from growing but also from melting. This melting inhibition by AFPs, which results in superheated ice (Celik et al, PNAS 2010), is not a well-known phenomenon. Here we present our recent findings in which the $Ca^{2+}$-dependent hyperactive AFP from *Marinomonas primoryensis* (*Mp*AFP) clearly displays this property. Additionally, we found that an ice crystal that is initially stabilized and protected by this type of AFP can be overgrown and then melted back to the original crystal. This repeatable process is likely due to melting inhibition, and supports the idea that AFPs bind irreversibly to ice surfaces. [1]

## 1. INTRODUCTION

Antifreeze proteins evolved in cold-adapted organisms and are well known for their ability to depress the freezing point, which leads to the protection of these organisms against freezing conditions[2]. AFPs are classified as moderate and hyperactive according to their freezing hysteresis activity. Freezing hysteresis is defined as the difference between the equilibrium melting point and the non-equilibrium freezing point, which is the temperature at which a crystal that is not growing in a supercooled solution suddenly grows rapidly (bursts). Although it has been more than 40 years since AFPs were discovered in Antarctic fish, how these proteins function at the ice-water interface remains a matter of debate. There is no consensus in the AFP field in regards to whether AFPs bind reversibly or irreversibly to ice surfaces. The adsorption inhibition model, which suggests that AFPs bind to ice surfaces irreversibly, was first introduced by Raymond *et al.* and further developed by Knight *et al.*[3,4]. The adsorption inhibition model indicates that ice will grow in the gaps where no AFPs are located[4]. This process increases the curvature of the ice surface between bound AFP molecules, thereby decreasing the radius of curvature from infinity to a finite magnitude. Such an increase in surface curvature leads to a depression of the freezing point due to Gibbs-Thomson effect, which states that the equilibrium melting point of a solid is related to the curvature of the particles and interfacial energy[2,4]. The absence of melting at temperatures higher than the equilibrium melting temperature is defined as the superheating of solids[5]. Knight and DeVries showed that antifreeze glycoproteins (AFGPs) can inhibit melting and concluded that ice in the presence of AFGPs can be superheated[6]. Although they were not able to show quantitatively that ice was superheated, those experiments were the first to show that AFPs can indeed prevent ice from melting. They suggested that the inhibition of melting can be explained by the same mechanism as that for the inhibition of freezing with a negative curvature between bound AFPs. In this case, the ice melts between the adsorbed proteins, which reduces the volume of ice and increases the area of water-ice interface. This increases the melting point of the concave ice. It was suggested that the amount of melting suppression should be comparable to the amount of freezing suppression[6]. However, until now there was no quantitative evidence of melting hysteresis.

Recently, we reported the first quantitative observations of the superheating of ice in several hyperactive and moderate AFP solutions[1]. We have found that the measured melting hysteresis was only around one tenth of the measured freezing hysteresis. Here, we provide additional evidence for the melting hysteresis phenomenon. We used wild-type *Mp*AFP to further describe Raman spectroscopy experiments of superheated ice crystals and also fluorescently tagged *Mp*AFP to visualize the adsorbed AFPs. In addition, we found that it is possible to repeatedly melt a crystal back to its original shape, after it has been overgrown by ice (burst) by successive cycles of cooling and warming. This repetitive phenomenon is of great importance in understanding the nucleation of crystals in general.

## 2. MATERIALS AND METHODS

### 2.1 Nanoliter osmometer experiments

In order to visualize micron-sized ice crystals, we used a custom designed nanoliter osmometer connected to bright field microscope (Olympus BH2)[1,7,8]. The cell consisted of a metal plate placed on top of a two-stage thermoelectric cooler (Laird Technologies, Melcor) in conjunction with a thermistor (GE Thermometric, Saint Mary's, PA). The system was driven by a temperature controller (Model 3140, Newport, Irvine, CA). The microscope was equipped with Nikon Air 50x (NA 0.55), Nikon Air 10x (NA 0.25), and Nikon Air 4x (NA 0.13) LWD objectives. The images were recorded with a Sony CCD-IRIS video camera and directed to a computer hard drive by using a video frame grabber device (IMAQ PCI-1407, National Instruments Inc., Austin, TX). The whole experimental system was controlled through Labview software. This temperature-controlled system allowed us to work with temperatures ranging from -40 $^{\circ}$C to room temperature with a precision of 0.002 $^{\circ}$C. Freezing hysteresis and melting hysteresis experiments were performed using this custom-designed nanoliter osmometer and these experiments have been described previously in detail[1]. In summary, a small amount of protein solution (~nl) was frozen by setting the temperature controller to -40 $^{\circ}$C. The sample typically froze at approximately -30 $^{\circ}$C. After the sample was frozen, the temperature was increased and the sample was melted continuously until only a single ice crystal with a diameter of 10-30 μm remained. The equilibrium melting point was defined as the temperature where the crystal melting velocity became imperceptible. Once the crystal of desired size was formed, the temperature of the sample was set 0.01 $^{\circ}$C to 0.3 $^{\circ}$C below the equilibrium melting temperature depending on the activity of the protein or concentration. The sample was held at this incubation temperature for 10 minutes. Unless otherwise described, the temperature was lowered 0.01 $^{\circ}$C every 4 seconds after the incubation period. The temperature at which rapid ice crystal growth (burst) occurred was defined as the non-equilibrium freezing temperature. Freezing hysteresis activity was defined as the difference between non-equilibrium freezing temperature and the equilibrium melting temperature.

A similar procedure was used to measure melting hysteresis activity[1]. The ice crystal was incubated for 10 minutes at 0.3 $^{\circ}$C below the equilibrium melting point. At the end of the incubation period, the temperature of the crystal was slowly increased at a rate of 0.01 $^{\circ}$C every 30 seconds. When the temperature crossed the equilibrium melting temperature the crystal did not melt. We observed a sudden melting of the ice crystal at higher temperatures with higher melting velocity. The difference between the temperature at which the superheated crystal actually melted and the equilibrium melting point was defined as the melting hysteresis.

### 2.2 Fluorescence microscopy experiments

We used an upright confocal microscope (Zeiss LSM 510, Thornwood, NY) to visualize fluorescently tagged AFPs[7,9,10]. The confocal microscope was equipped with 488 nm and 633 nm illumination lines and filters for the detection of GFP and Cyanine 5 (Cy5), respectively. The experimental cell used in this setup was basically the same as that used in the nanoliter osmometer setup except that the AFP solution was sandwiched in between two cover glasses. The base cover glass was a square of 22 mm x 22 mm in size, whereas the upper cover glass was circular with a diameter of 18 mm. The peripheral area of the upper cover glass was sealed with a non-cured silicone elastomer, Polydimethylsiloxane (PDMS) (Slygard 184, Dow Corning Corp., Midland, MI). The sandwiched cover glasses were placed on a metal plate to control its temperature and imaging was conducted through 125 μm diameter holes in the metal plate. Cy5 was used to increase the contrast between proteins adsorbed on the ice surface and the proteins in solution[10].

## 2.3 Raman spectroscopy experiments[1]

The temperature-controlled cell was placed under a WiTec Raman/near-field scanning optical microscope (WiTec Instruments Corp., Maryville, TN), which had an illumination line of 532 nm. We used a Nikon Air 50x (NA 0.55) LWD objective to collect Raman spectra of ice and solution phases. As in the fluorescence microscopy experiments, we sandwiched a sample (5 μl) in between two cover glasses which were placed on a temperature-controlled metal plate. The sample was frozen and melted back to form individual ice crystals. Raman images as well as single spectra and time spectra of water, ice below melting point, and ice in the superheated state were collected[1].

## 3. RESULTS AND DISCUSSION

In Figure 1, an ice crystal in a solution of 10 μM GFP-*Mp*AFP is shown. Once the equilibrium melting point was measured as -0.18 °C, the crystal was stabilized at -0.48 °C for 10 minutes. The temperature was increased by a rate of 0.01 °C per 30 seconds. Once the temperature reached -0.05 °C, the crystal melted rapidly. Note that this specific crystal was held 0.13 °C above equilibrium melting point for 7 seconds before melting, and then the crystal melted in 0.23 seconds with a velocity of 62.5 μm/s (Fig.1).

In addition to this experiment, we repeated an experiment done previously using *Mp*AFP at 36 μM [1] with a lower concentration (6.4 μM). Freezing and melting hysteresis measurements were done with the same crystal. First we measured the freezing hysteresis activity, which, as expected, was low at ~0.02 °C. When the sample was melted back, the ice disappeared at a steady rate until only the initial ice crystal remained, at which point the melting ceased. When we cooled the crystal again, it burst at different positions each time (Fig. 2). This behavior indicates that the nucleation of new ice on the crystal (The position on the crystal where the growth inhibition by AFPs failed) is random and does not occur at specific sites. We intend to investigate this phenomenon further but did observe that the bursts consistently occurred on the flat region of the crystal. From our previous studies, we know that the flat regions possess less AFPs than the corners[7]. The mode of growth, which is rotated 30° compared to the hexagonal orientation of the original crystal, is consistent with the asymmetry between growth and melting that we reported previously[7]. This asymmetry of growth and melt was also observed in other systems such as ice crystals under high pressure and low temperature[11]. The results presented in Figure 2 demonstrate that the hexagonal shape was formed during melting (Fig 2. A, F, & L) and when growth occurs, the flat regions developed into corners (Fig 2. B, C, G, & H). Our experiments clearly show that ice was protected by AFPs, and that melting hysteresis is evident even with a low concentration of AFP.

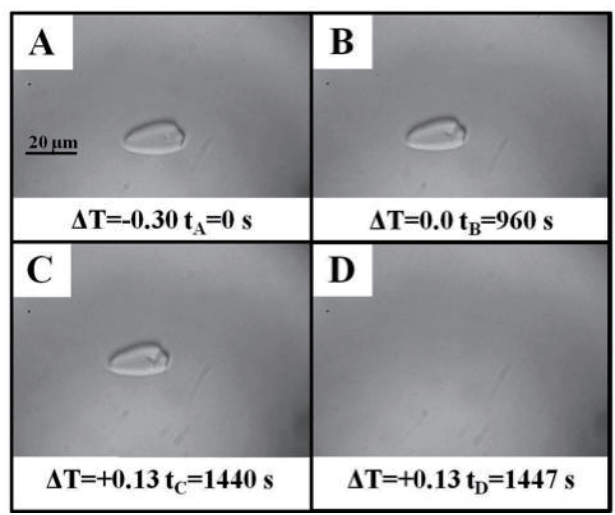

Figure 1. **Superheated ice in *Mp*AFP solution.** (a) A single ice crystal was grown in a 10 μM GFP-*Mp*AFP solution and stabilized for 10 minutes at 0.3 °C below its equilibrium melting temperature. (b) When brought back to its equilibrium melting temperature, the crystal remained intact. (c) The crystal was superheated and was not melting at 0.13 °C above its melting temperature. (d) After remaining stable for nine seconds at this superheated temperature, the ice crystal melted rapidly.

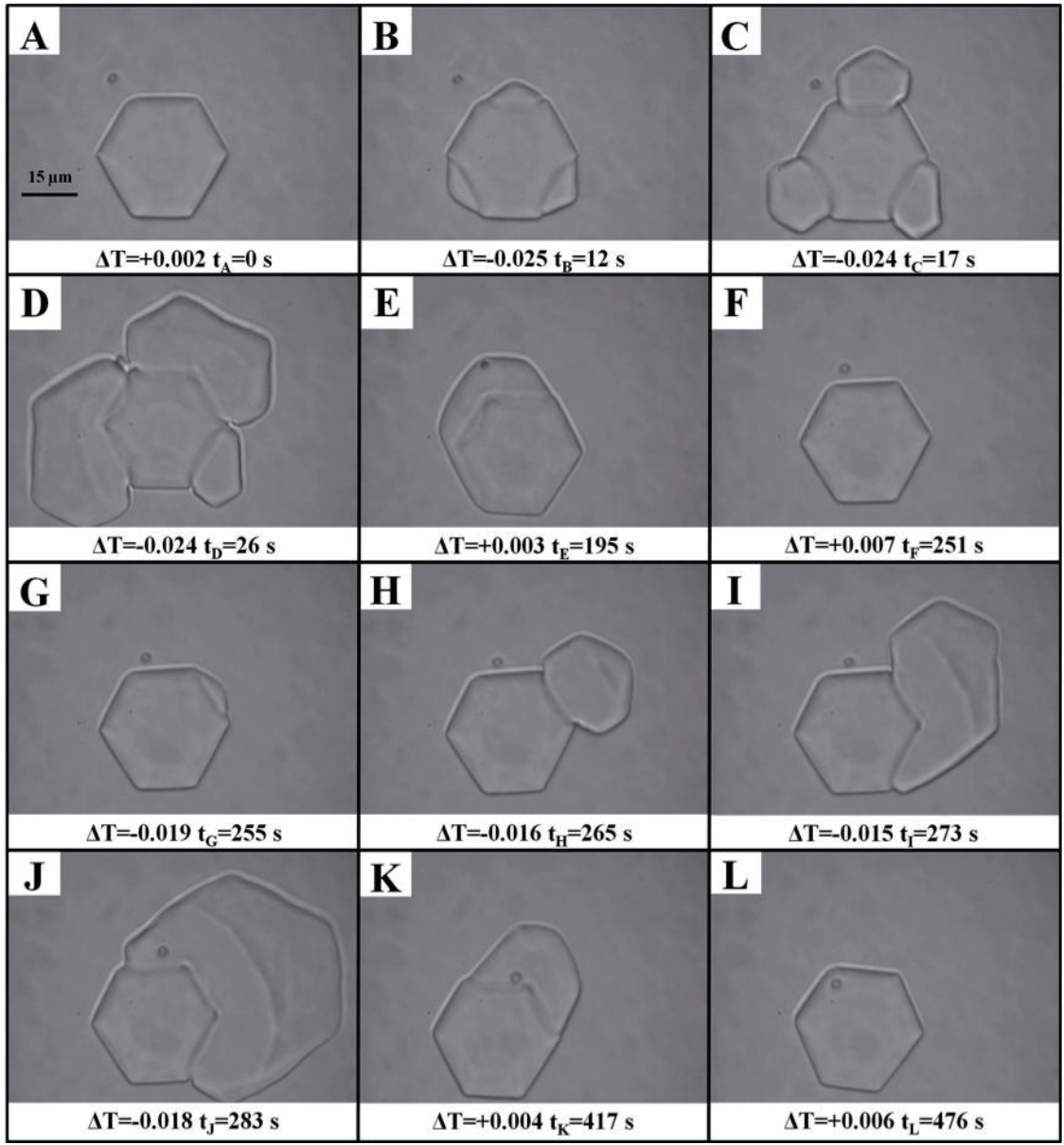

Figure 2. **Superheated ice crystal in low concentrations of *Mp*AFP solution.** A single ice crystal was formed by melting back a frozen droplet of 6.4 μM *Mp*AFP solution. The crystal was warmed and slightly superheated (A), then it was cooled to 0.025 °C below the melting point at which point it burst from 3 different faces independently (B) and continued to grow (C-D). When the crystal was warmed above the melting point, the ice melted back then stopped melting at the edge of the original crystal (E). The crystal returned to the original shape and was stable when superheated by 0.007 °C (F). When cooled again, the crystal burst from a single face (G), and continued to grow (H-J). When warmed again (K-L), the same phenomenon was observed as in E-F.

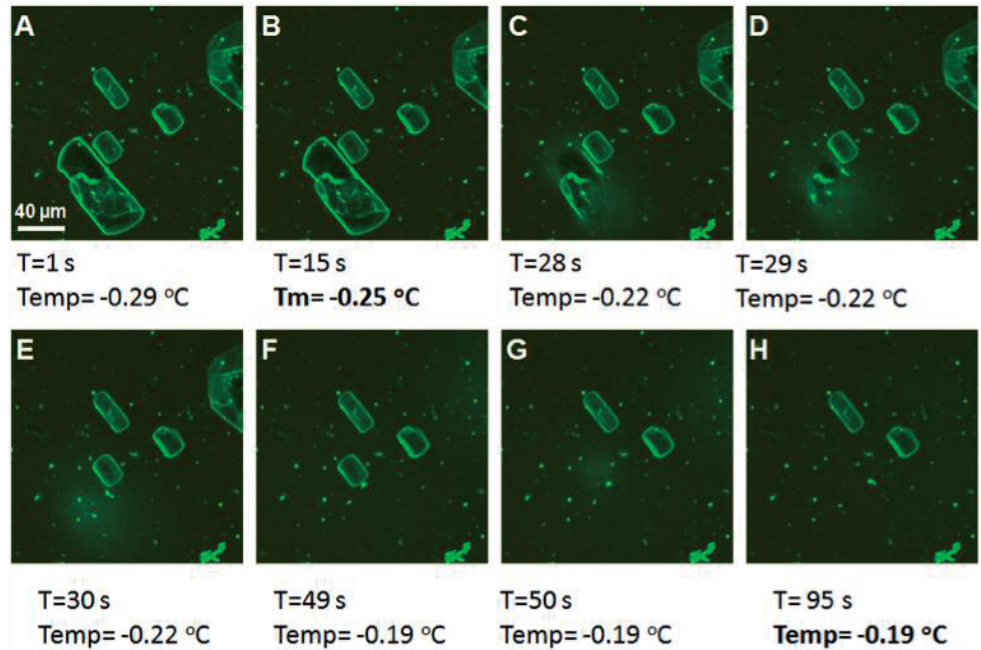

Figure 3. **Ice crystals in fluorescently tagged *Mp*AFP solution.** Ice crystals in a solution of 16 μM GFP-*Mp*AFP imaged with confocal microscopy (A). Crystals were stable at the equilibrium melting point ($T_m$) (B). As the temperature was increased slowly (C-F), some crystals started to melt. Only two crystals (imaged at different times) were left at 0.06 °C above the equilibrium melting temperature (G-H).

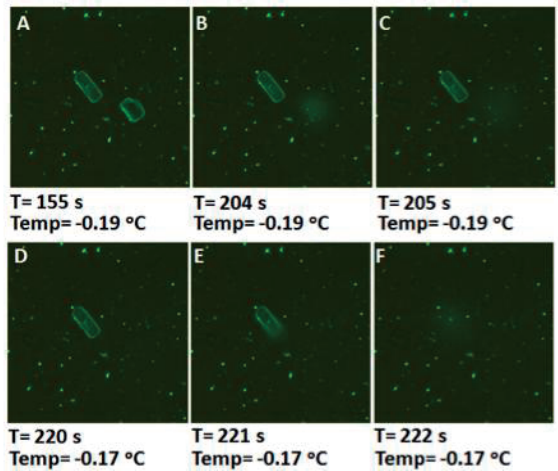

Figure 4. **Superheated ice crystals in GFP-*Mp*AFP solution.** This is a continuation of the experiment described in Figure 3. The equilibrium melting temperature ($T_m$) of this sample was -0.25 °C; these two crystals were already superheated to 0.06 °C above $T_m$ (A).After some time the crystal on the lower right corner melted and the proteins adsorbed on the surface diffused away (B-C). When we further increased the temperature, the last remaining crystal melted at 0.08 °C above the $T_m$ (D-F).

Superheating of ice crystals was also confirmed with fluorescence microcopy experiments[1]. The GFP-*Mp*AFP solution (16 μM) was sandwiched between two cover glasses, frozen at -20 °C then brought back to melting temperatures. The equilibrium melting temperature of this sample was measured as -0.25 °C. As the temperature was increased slowly, the crystals started to melt, one by one. When the temperature reached -0.19 °C, there were only two crystals remaining. As shown in Fig. 4, the last crystal melted once the temperature increased to 0.08 °C above the equilibrium melting temperature. The proteins that were adsorbed to ice surfaces diffused away as the crystals melted. These results are consistent with our previous findings in which, using the nanoliter osmometer device, a group of ice crystals was observed for several hours as the temperature was increased slowly[1]. Here too, there seems to be an inverse correlation between crystal size and superheating. However, the correlation is weak as we observed that some small crystals melted earlier than the big ones. Nonetheless, each experiment should be considered carefully as the time of the formation of each crystal plays a significant role. Crystals formed earlier in the process are protected more as they have had more time for proteins to adsorb on their surfaces, whereas the crystals formed later in the process are less protected and tend to melt first.

It is possible to differentiate ice from water as well as to distinguish different ice forms by their Raman O-H stretch band signature[12]. We addressed the question of whether superheated ice is different from non-superheated ice by recording the Raman spectrum of crystals below and above the equilibrium melting point. The crystal shown (Fig. 5) was grown in 72 µM of *Mp*AFP solution and incubated for 45 minutes at 0.05 °C below its melting point. The temperature was then slowly increased stepwise while images and single spectra were taken continuously between temperature increments. The series of images in Figure 5A shows this crystal below the melting temperature and at two different superheating temperatures. Figure 5B shows a Raman intensity map of a 30 µm x 30 µm section of this ice crystal before and after it was superheated[1]. The maximum measured melting hysteresis for this particular crystal was 0.37 °C[1]. The Raman spectra of the ice were collected from the center of the crystal at different temperatures in order to observe the spectral change as it went through phase transitions at the limit of superheating (Fig. 5C). The local maximum (around 640 nm) that appeared in the Raman spectrum of the crystal both before and after it was superheated, but not in the neighboring water, clearly indicated that indeed we have superheated ice and it is still in the ice Ih structure[13,1]. The Raman spectra of ice observed in these experiments are in agreement with other spectral analyses of ice Ih[12,13].

## 4. SUMMARY

In conclusion, we have further demonstrated our previous findings[1] that ice can be superheated in AFP solutions and that AFPs irreversibly bind to ice surfaces. We showed that ice crystals that are protected by AFP can be recognized as type Ih ice by Raman spectrometry and that the ice maintains its structural integrity when superheated[1]. We also showed that melting inhibition by AFPs enables us to observe the repeatable nucleation process in the same crystal and the burst of ice crystals in aqueous solutions. This phenomenon opens the possibility to further investigate the nucleation process. The results of this study are of significant importance to the current understanding of the interaction of AFPs, in particular hyperactive AFPs, with ice crystals.

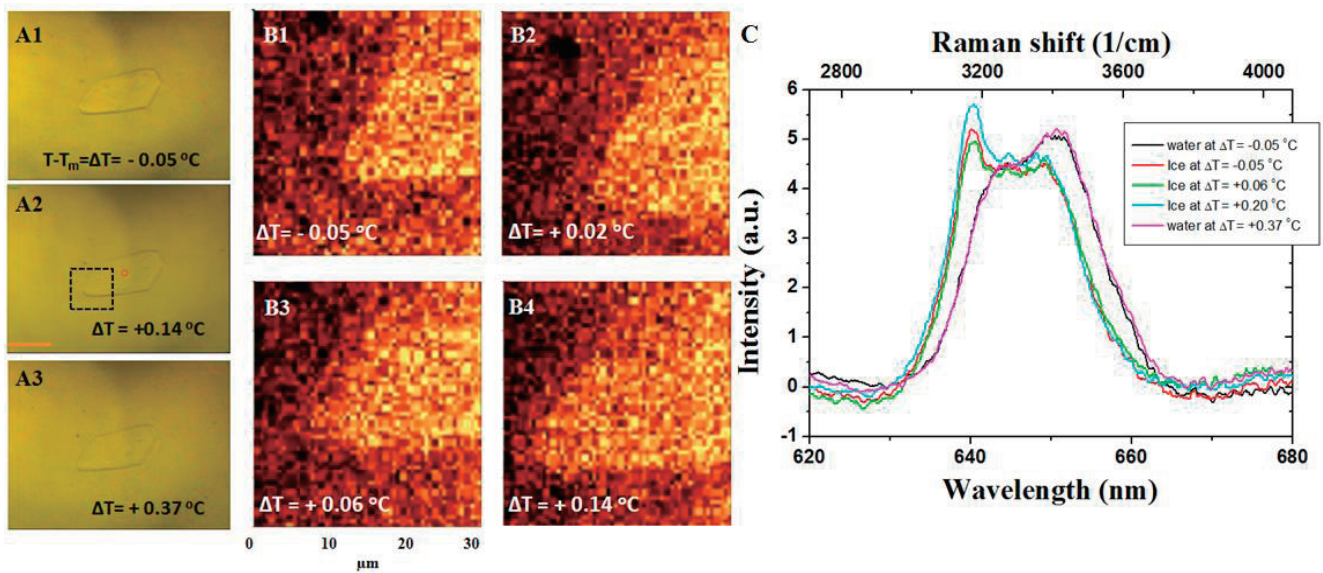

Figure 5. **Raman Spectra of ice, superheated ice, and water[1].** Ice crystal at different temperatures is shown in series of images (A1-A3). The dotted rectangle indicates the area used to obtain a Raman intensity map of a 30 µm x 30 µm section of the ice crystal (B1-B4). The red circular dot spot in the center of the crystal indicates where the laser is pointed to collect the Raman spectrum from the ice. (C) The Raman spectra obtained for water and ice at different temperatures. The pink line shows the spectrum of the solution right after the ice melted at the same spot that the ice spectra was taken. Adapted from supplementary online materials of reference 1.

## ACKNOWLEDGMENTS

The authors are grateful to S. Gauthier, C. Garnham, and J. Whitney for their assistance in preparing AFPs, to Prof. H. Richardson and A. Khan for their help with Raman spectroscopy and to Tyler Barton for his help in editing the manuscript. This work has been supported by the National Science Foundation (NSF) under Grant No. CHE-0848081, the Israel Science Foundation (grant No. 1279/10 and 1281/10), by Marie Curie International Reintegration Grant (Grant No. 256364), the Canadian Institutes for Health Research (CIHR), the Condensed Matter and Surface Science program at Ohio University (CMSS), and the Biomimetic Nanoscience and Nanoscale Technology initiative (BNNT) at Ohio University.